\documentclass[apj]{emulateapj}
\newcommand\submitms{n}		
\if\submitms y
\else
\usepackage{apjfonts}
\fi

\usepackage{ifthen}
\usepackage{natbib}
\usepackage{amssymb, amsmath}
\usepackage{appendix}

\if\submitms y
\bibpunct[, ]{(}{)}{,}{a}{}{,}
\else
\fi

\shorttitle{Thermal Emission of WASP-43\MakeLowercase{b}}
\shortauthors{Blecic {\em et al.}}

\newcommand\degree{\degr}
\newcommand\degrees\degree

\DeclareSymbolFont{UPM}{U}{eur}{m}{n}
\DeclareMathSymbol{\umu}{0}{UPM}{"16}
\let\oldumu=\umu
\renewcommand\umu{\ifmmode\oldumu\else\math{\oldumu}\fi}
\newcommand\micro{\umu}
\renewcommand\micron{\micro m}
\newcommand\microns \micron

\let\oldsim=\sim
\renewcommand\sim{\ifmmode\oldsim\else\math{\oldsim}\fi}
\let\oldpm=\pm
\renewcommand\pm{\ifmmode\oldpm\else\math{\oldpm}\fi}
\newcommand\by{\ifmmode\times\else\math{\times}\fi}
\newcommand\ttt[1]{10\sp{#1}}
\newcommand\tttt[1]{\by\ttt{#1}}
\newcommand\tablebox[1]{\begin{tabular}[t]{@{}l@{}}#1\end{tabular}}
\newbox{\wdbox}
\renewcommand\c{\setbox\wdbox=\hbox{,}\hspace{\wd\wdbox}}
\renewcommand\i{\setbox\wdbox=\hbox{i}\hspace{\wd\wdbox}}

\newcount\timect
\newcount\hourct
\newcount\minct
\newcommand\now{\timect=\time \divide\timect by 60
         \hourct=\timect \multiply\hourct by 60
         \minct=\time \advance\minct by -\hourct
         \number\timect:\ifnum \minct < 10 0\fi\number\minct}

\newcommand\mctc{\multicolumn{2}{c}}

\catcode`@=11

\if\submitms y
\renewcommand\comment[1]{}
\else
\newcommand\comment[1]{}
\fi
\newcommand\commenton{\catcode`\%=14}
\newcommand\commentoff{\catcode`\%=12}

\renewcommand\math[1]{$#1$}
\newcommand\mathshifton{\catcode`\$=3}
\newcommand\mathshiftoff{\catcode`\$=12}

\comment{the backslash is necessary}

\comment{alignment tab}

\let\atab=&
\newcommand\atabon{\catcode`\&=4}
\newcommand\ataboff{\catcode`\&=12}

\let\oldmsp=\sp
\let\oldmsb=\sb
\def\sp#1{\ifmmode
           \oldmsp{#1}%
         \else\strut\raise.85ex\hbox{\scriptsize #1}\fi}
\def\sb#1{\ifmmode
           \oldmsb{#1}%
         \else\strut\raise-.54ex\hbox{\scriptsize #1}\fi}
\newbox\@sp
\newbox\@sb
\def\sbp#1#2{\ifmmode%
           \oldmsb{#1}\oldmsp{#2}%
         \else
           \setbox\@sb=\hbox{\sb{#1}}%
           \setbox\@sp=\hbox{\sp{#2}}%
           \rlap{\copy\@sb}\copy\@sp
           \ifdim \wd\@sb >\wd\@sp
             \hskip -\wd\@sp \hskip \wd\@sb
           \fi
        \fi}
\def\msp#1{\ifmmode
           \oldmsp{#1}
         \else \math{\oldmsp{#1}}\fi}
\def\msb#1{\ifmmode
           \oldmsb{#1}
         \else \math{\oldmsb{#1}}\fi}
\def\supon{\catcode`\^=7}
\def\supoff{\catcode`\^=12}
\def\subon{\catcode`\_=8}
\def\suboff{\catcode`\_=12}
\def\supsubon{\supon \subon}
\def\supsuboff{\supoff \suboff}

\comment{
\let\oldmsp=\sp
\let\oldmsb=\sb
\renewcommand\sp[1]{\ifmmode
	   \oldmsp{#1}%
	 \else\strut\raise.85ex\hbox{\scriptsize #1}\fi}
\renewcommand\sb[1]{\ifmmode
	   \oldmsb{#1}%
	 \else\strut\raise-.54ex\hbox{\scriptsize #1}\fi}
\newcommand\msp[1]{\ifmmode
	   \oldmsp{#1}
	 \else \math{\oldmsp{#1}}\fi}
\newcommand\msb[1]{\ifmmode
	   \oldmsb{#1}
	 \else \math{\oldmsb{#1}}\fi}
\newcommand\supon{\catcode`\^=7}
\newcommand\supoff{\catcode`\^=12}
\newcommand\subon{\catcode`\_=8}
\newcommand\suboff{\catcode`\_=12}
\newcommand\supsubon{\supon \subon}
\newcommand\supsuboff{\supoff \suboff}
}

\newcommand\actcharon{\catcode`\~=13}
\newcommand\actcharoff{\catcode`\~=12}

\newcommand\paramon{\catcode`\#=6}
\newcommand\paramoff{\catcode`\#=12}

\comment{And now to turn us totally on and off...}

\newcommand\reservedcharson{\commenton \mathshifton \atabon \supsubon \actcharon
	\paramon}

\newcommand\reservedcharsoff{\commentoff \mathshiftoff \ataboff
	\supsuboff \actcharoff \paramoff}

\catcode`@=12
\reservedcharsoff

\reservedcharson

\comment{ Must have ONLY ONE of these... trust these macros, they work

}
\reservedcharsoff

\actcharon
\if\submitms y

\else

\comment{
\received{}
\revised{}
\accepted{}
}
\fi

\if\submitms n
\slugcomment{\tablebox{
doi:10.1088/0004-637X/781/2/116. Published 2014 January 16 in {\em ApJ} . \\
}}
\else
\slugcomment{Submitted to {\em ApJ}. November 8, 2011 15:26}
\journalinfo{}
\fi

\reservedcharson
\begin{document}




\title {{\em Spitzer} Observations of the Thermal Emission from WASP-43b}

\author{Jasmina Blecic\altaffilmark{1},      Joseph Harrington\altaffilmark{1, 2},
        Nikku Madhusudhan\altaffilmark{3},   Kevin B.\ Stevenson\altaffilmark{1},
        Ryan A.\ Hardy\altaffilmark{1},      Patricio E.\ Cubillos\altaffilmark{1, 2},
        Matthew Hardin\altaffilmark{1},      Oliver Bowman \altaffilmark{1},
        Sarah Nymeyer\altaffilmark{1},
        David R.\ Anderson\altaffilmark{4},  Coel Hellier\altaffilmark{4},
        Alexis M.\ S.\ Smith\altaffilmark{4} and  Andrew Collier Cameron\altaffilmark{5}}

\affil{\sp1 Planetary Sciences Group, Department of Physics, University of Central Florida, Orlando, FL 32816-2385, USA}

\affil{\sp2 Max-Planck-Institut f\"{u}r Astronomie, D-69117 Heidelberg, Germany}

\affil{\sp3 Department of Physics and Department of Astronomy, Yale University, New Haven, CT 06511, USA}

\affil{\sp4 Astrophysics Group, Keele University, Keele, Staffordshire ST5 5BG, UK}

\affil{\sp5 SUPA, School of Physics and Astronomy, University of St.\ Andrews, North Haugh, St.\ Andrews, Fife KY16 9SS, UK}

\email{jasmina@physics.ucf.edu}

\begin{abstract}

WASP-43b is one of the closest-orbiting hot Jupiters, with a semimajor axis of \math{a} = 0.01526 {\pm} 0.00018 AU and a period of only 0.81 days. However, it orbits one of the coolest stars with a hot Jupiter (\math{T\sb{\rm{*}}} = 4520 {\pm} 120 K), giving the planet a modest equilibrium temperature of \math{T\sb{\rm{eq}}} = 1440 {\pm} 40 K, assuming zero Bond albedo and uniform planetary energy redistribution. The eclipse depths and brightness temperatures from our jointly fit model are 0.347\% {\pm} 0.013\% and 1670 {\pm} 23 K at 3.6 {\micron} and 0.382\% {\pm} 0.015\% and 1514 {\pm} 25 K at 4.5 {\micron}. The eclipse timings improved the estimate of the orbital period, \math{P}, by a factor of three (\math{P} = 0.81347436 {\pm} 1.4\tttt{-7} days) and put an upper limit on the eccentricity (\math{e = 0.010^{+0.010}_{-0.007}}). We use our {\em Spitzer}\/ eclipse depths along with four previously reported ground-based photometric observations in the near-infrared to constrain the atmospheric properties of WASP-43b. The data rule out a strong thermal inversion in the dayside atmosphere of WASP-43b. Model atmospheres with no thermal inversions and fiducial oxygen-rich compositions are able to explain all the available data. However, a wide range of metallicities and C/O ratios can explain the data. The data suggest low day-night energy redistribution in the planet, consistent with previous studies, with a nominal upper limit of about 35\% for the fraction of energy incident on the dayside that is redistributed to the nightside.

\if\submitms y
\else
\fi
\end{abstract}

\section{INTRODUCTION}
\label{intro}

Our knowledge of exoplanetary systems is rapidly improving.
Recent {\em Kepler}\/ results \citep{Borucki2011ApKeplerFirstResults,Batalha2012Kepler-secondResults} have shown a striking increase in detections of the smallest candidates, and the planet candidate lists now show that hot Jupiters are much less common than planets smaller than Neptune.  However, nearly all {\em Kepler}\/ candidates are too small, cold, or distant for atmospheric characterization, except the nearby hot Jupiters. Their host stars, bright enough for radial velocity (RV) measurements, subject these planets to a strong irradiating flux, which governs their atmospheric chemistry and dynamics. Their large sizes and large scale heights \citep[e.g.,][]{ShowmanGuillot2002A&A} give the signal-to-noise ratio (S/N) needed for basic atmospheric characterization.

The most common technique for observing hot Jupiters and characterizing their dayside atmospheres is secondary eclipse photometry \citep[e.g., ][]{FraineEtal2013-GJ1214b, CrossfieldEatl2012ApJWASP12b-reavaluation, TodorovEtal2012ApJ-XO4b-HATP6b-HATP8b, DesertEatl2012AAS-Spizter, DemingEtal2011ApJ-CoRoT1-CoRoT2, BeererEtal2011ApJ-WASP4b, Demory2007aaGJ436bspitzer}. During secondary eclipse, when the planet passes behind its star, we see a dip in integrated flux proportional to the planet-to-star flux ratio, or usually 0.02\%--0.5\% in the {\em Spitzer Space Telescope} infrared wavelengths, where the signal is strongest.  This dip is much lower at wavelengths accessible from the ground or from the {\em Hubble Space Telescope}. Techniques such as phase curve measurement \citep{KnutsonEatl2009ApJ-PhaseVariationHD149026b, Knutson2012ApJ-phaseVariation-HD189733b, LewisEtal2013arXiv-PhaseVAr-HATP2b, CowanEtal2012ApJ-ThermalPhaseVar-WASP12b, CowanEtal2012ApJ-ThermalPhases-EarthLikePlanets, CrossfieldEtal2010apjUpsAndb}, transmission spectroscopy \citep{DemingEtal2013arXivHD209458b-Xo1b-WCF3, GibsonEtal2012MNRAS-HD189733b-WFC3, BertaEtal2012ApJ-GJ1214b-WFC3}, and ingress--egress mapping \citep{deWitEtal2012aapFacemap, MajeauEtal2012Facemap} can reveal more than a secondary eclipse but are available for only a small number of high-S/N planets.

A secondary eclipse observed in one bandpass places a weak constraint on an exoplanet's temperature near the average altitude of optical depth unity over that bandpass. Multiple wavelengths constrain the planet's dayside spectrum, potentially yielding insight into the atmospheric composition and temperature structure. Different wavelengths probe different atmospheric levels and can be combined into a broadband spectrum for further atmospheric modeling \citep[e.g.,][]{MadhusudhanSeager2009, StevensonEtal2010Natur}.  Infrared observations are specifically valuable because the most abundant chemical species in planetary atmospheres (aside from H\sb{2} and He), such as H\sb{2}O, CO, CO\sb{2}, and CH\sb{4}, have significant absorption and emission features at these wavelengths \citep[e.g.,][]{MadhusudhanSeager2010}. Constraints on chemical composition and thermal structure are important for both further atmospheric modeling \citep[e.g.,][]{ShowmanEtal2009ApJ-3DCircModel} and studies of the planet's formation. Several recent studies have shown that the atmospheric C/O ratios of giant planets can be significantly different from those of their host stars because of, for example, the formation location of the planet (see, for exoplanets, \citealp{Obergetal2011ApJ-C/O} and \citealp{Madhusudhan2011b}, and for Jupiter, \citealp{Lodders2004ApJ-Jupiter} and \citealp{MousisEtal2012-Jupiter}).

Secondary eclipse observations also provide insight into the exoplanet's orbit. Measuring the time of the secondary eclipse relative to the time of transit can establish an upper limit on orbital eccentricity, \math{e}, and constrain the argument of periapsis, \math{\omega}, independently of RV measurements. Orbital eccentricity is important in dynamical studies and in calculating irradiation levels.  Apsidal precession can also be constrained by eclipse timing and can be used to reveal the degree of central concentration of mass in the planetary interior \citep{ragozzine:2009, CampoEtal2011apjWASP12b, LopezMoralesEtal2010apjWASP12borb}.

WASP-43b was first detected by the Wide-Angle Search for Planets (WASP) team \citep{Hellier2011-WASP43b} in 2009 and 2010 from the WASP-South and WASP-North observatories. The WASP team also performed follow-up measurements with the CORALIE spectrograph, the TRAPPIST telescope, and EulerCAM in 2010 December. These observations revealed a planet with a mass of \math{M\sb{\rm p}} = 1.78 Jupiter masses (\math{M\sb{\rm{J}}}) and a radius of \math{R\sb{\rm p}} = 0.93 Jupiter radii (\math{R\sb{\rm J}}), transiting one of the coldest stars to host a hot Jupiter (type K7V, \math{T\sb{*}} = 4400 {\pm} 200 K). They found the planet to have an exceptionally short orbital period of 0.81 days and a semimajor axis of only 0.0142 AU, assuming the host star has a mass of \math{M\sb{*}} = 0.58 {\pm} 0.05 \math{M\sb{\odot}}. The planet's orbital eccentricity was constrained by the radial velocity and transit data to \math{e} < 0.04 at 3\math{\sigma}.  Spectroscopic measurements of the star revealed a surface gravity of log\,\math{(g)} = 4.5 {\pm} 0.2 (cgs) and a projected stellar rotation velocity of \math{v\sb{*}\sin(i)} = 4.0 {\pm} 0.4 km\,s\sp{-1}, where \math{i} is the inclination of the star's pole to the line of sight. Strong Ca H and K emission indicates that the star is active. The estimated age of the star is \math{400^{+200}_{-100}} Myr.

For low-mass stars like WASP-43, there are notable discrepancies \citep{BergerEtal2006ApJ-LowMassStars} between interferometrically determined radii and radii calculated in evolutionary models \citep[i.e.,][]{ChabrierBaraffe1997-LowMassStars, SiessEtal1997-SyntheticHRDiagram}.  \citet{BoyajianEtal2012-KMStars} presented high-precision interferometric diameter measurements of 33 late-type K and M stars. They found that evolutionary models overpredict temperatures for stars with temperatures below 5000 K by \sim 3\% and underpredict radii for stars with radii below 0.7 \math{R\sb{\odot}} by \sim 5\%. Their Table 11 lists an average temperature and radius for each spectral type in the sample, suggesting that WASP-43, with its measured temperature of 4520 {\pm} 120 K, is likely a K4 star rather than a K7 as reported by \citet{Hellier2011-WASP43b}.

\citet{GillonEtal2012A&A-WASP-43b} analyzed twenty-three transit light curves, seven occultations, and eight new measurements of the star's RV, observed during 2010 and 2011 with TRAPPIST, the Very Large Telescope (VLT), and EulerCAM. They refined eccentricity to \math{e} = 0.0035 {\pm} 0.0043 and placed a 3\math{\sigma} upper limit of 0.0298 using all data simultaneously. They also improved the parameters of the system significantly (\math{M\sb{\rm{p}}} = 2.034 {\pm} 0.052 \math{M\sb{\rm{J}}}, \math{R\sb{\rm{p}}} = 1.036 {\pm} 0.019 \math{R\sb{\rm{J}}}), refined stellar parameters (\math{T\sb{\rm{eff}}} = 4520 {\pm} 120 K, \math{M\sb{*}} = 0.717 {\pm} 0.025 \math{M\sb{\odot}}, \math{R\sb{*}} = 0.667 {\pm} 0.011 \math{R\sb{\odot}}), and constrained stellar density (\math{\rho\sb{*}} = 2.41 {\pm} 0.08 \math{\rho\sb{\odot}}). They also confirmed that the observed variability of the transit parameters can be attributed to the variability of the star itself (consistent with \citealp{Hellier2011-WASP43b}). In addition, they detected the planet's thermal emission at 1.19 {\micron} and 2.09 {\micron} and used the atmospheric models of \citet{Fortney2005, Fortney2008} to infer poor redistribution of heat to the night side and an atmosphere without a thermal inversion.

In this paper we present two secondary eclipses, observed at 3.6 and 4.5 {\micron} with the Infrared Array Camera \citep[IRAC,][]{Fazio2004IRAC} on the {\em Spitzer Space Telescope}, which further constrain the dayside emission of the planet and improve the orbital parameters of the system. We combine our {\em Spitzer} eclipse depth measurements with on previously reported measurements of thermal emission in the near-infrared from \citet{GillonEtal2012A&A-WASP-43b} and \citet{WangEtal2013-WASP43b} to constrain the atmosphere's energy redistribution and thermal profile by using the retrieval method of \citet{MadhusudhanSeager2009} as subsequently developed.

The following sections present our observations (Section \ref{sec:obs}); discuss photometric analysis (Section \ref{sec:method}); explain specific steps taken to arrive at the fits for each observation and a joint fit (Section \ref{sec:results}); give improved constraints on the orbital parameters based on available RV, eclipse, and transit data (Section \ref{sec:orbit}); discuss implications for the planetary emission spectrum and planetary composition (Section \ref{sec:atm}); state our conclusions (Section \ref{sec:conc}); and, in the Appendix, supply the full set of system parameters from our own work and previous work.  The electronic attachment to this paper includes archival light curve files in FITS ASCII table and IRSA formats.

\section{OBSERVATIONS}
\label{sec:obs}

\begin{figure*}[ht]
\vspace{-5pt}
\strut\hfill
\includegraphics[width=0.30\textwidth, clip]{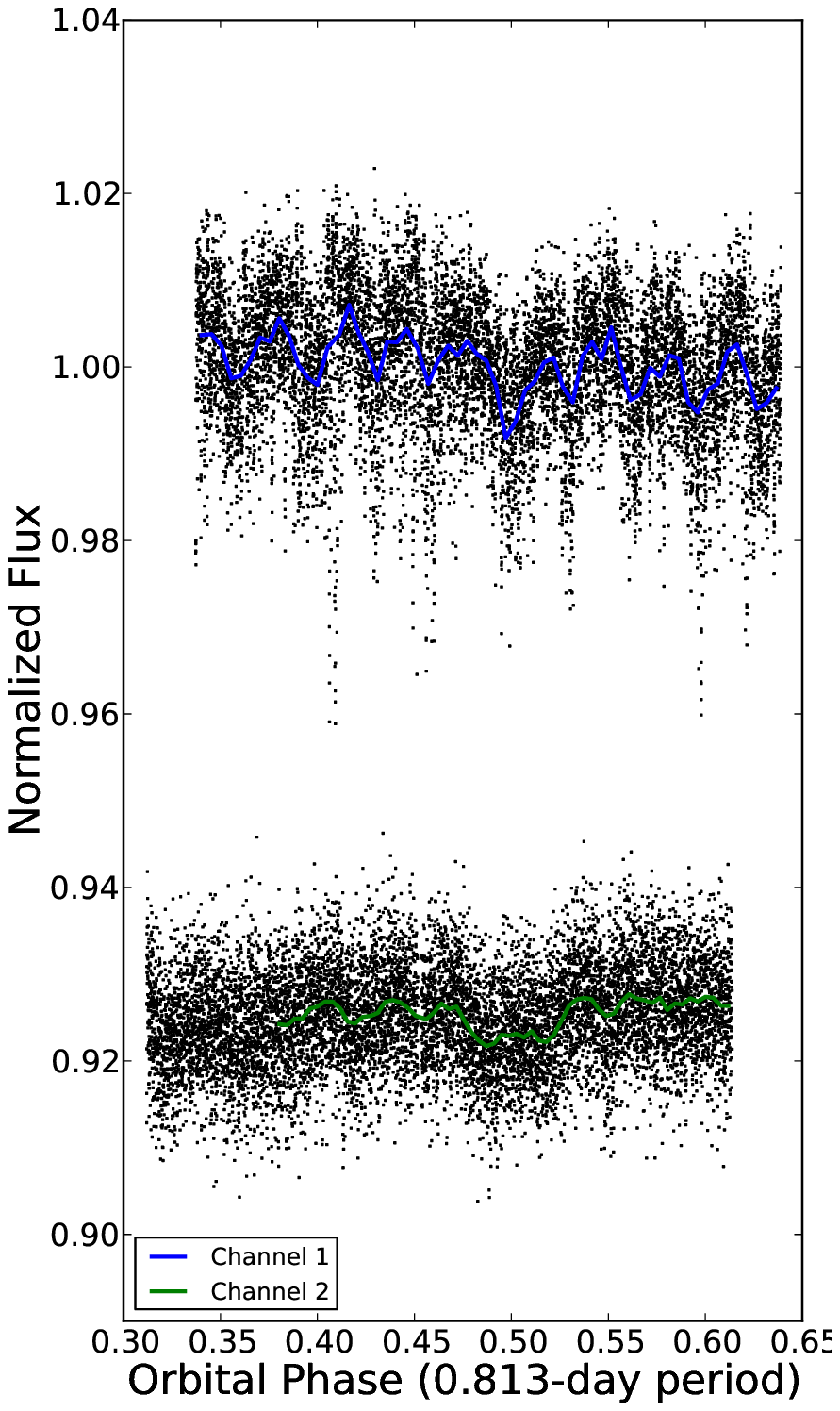}\hfill
\includegraphics[width=0.30\textwidth, clip]{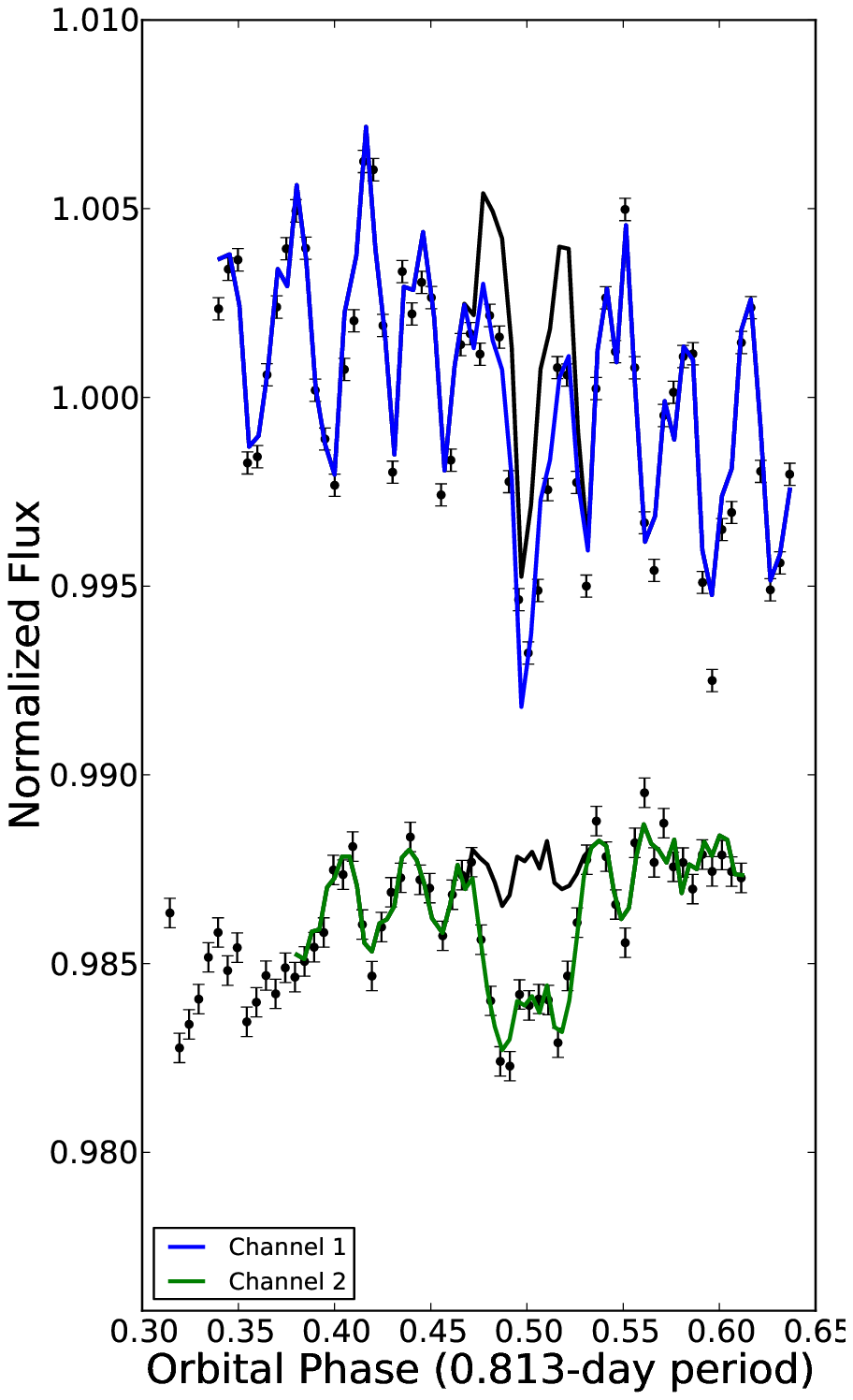}\hfill
\includegraphics[width=0.30\textwidth, clip]{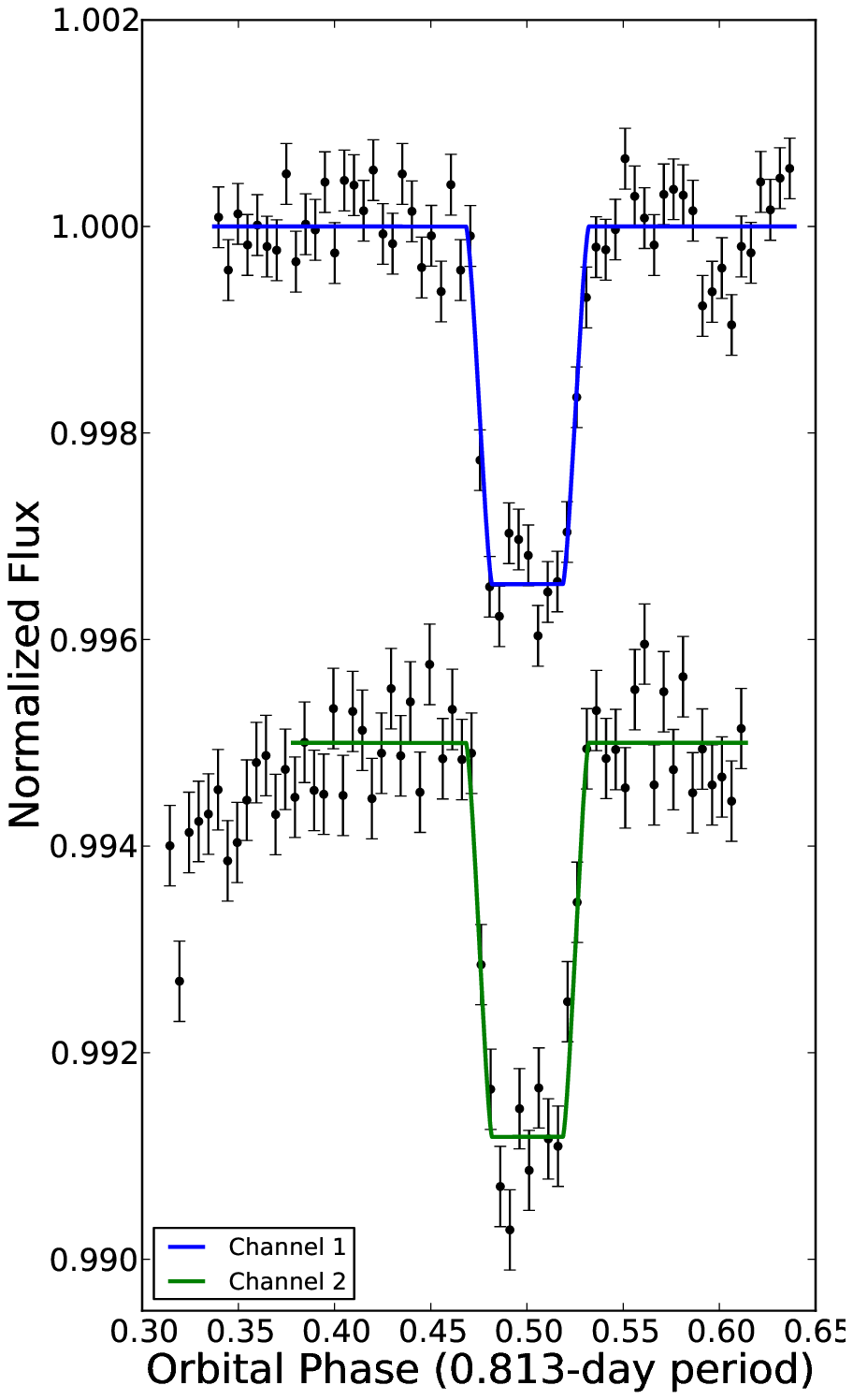}
\hfill\strut
\figcaption{\label{fig:RawBinNorm}
Raw (left), binned (center, 60 points per bin, with 1\math{\sigma} error bars), and systematics-corrected (right) secondary eclipse light curves of WASP-43b at 3.6 and 4.5 {\microns}. The results are normalized to the system flux and shifted vertically for comparison.  Note the different vertical scales used in each panel. The colored lines are best-fit models. The black curves in panel 2 are models without eclipses. As seen in the binned plots of channel 2, a ramp model is not needed to correct for the time-dependent systematic if initial data points affected by pointing drift are clipped (see Section \ref{sec:method}).
}
\end{figure*}

We observed two secondary eclipses of WASP-43b with the {\em Spitzer}\/ IRAC camera in subarray mode (Program ID 70084).  A sufficiently long baseline (Figure \ref{fig:RawBinNorm}) was monitored before the eclipses, providing good sampling of all {\em Spitzer}\/ systematics. To minimize intrapixel variability, each target had fixed pointing. We used the Basic Calibrated Data (BCD) from {\em Spitzer}'s data pipeline, version S.18.18.0. Basic observational information is given in Table \ref{table:ObsDates}.

\begin{table}[h]
\caption{\label{table:ObsDates} 
Observation Information}
\atabon\strut\hfill\begin{tabular}{lccc@{ }c@{ }c@{ }l}
    \hline
    \hline 
    Channel & Observation & Start Time    & Duration  & Exposure  & Number of \\
            & Date        & (MJD\sb\rm{UTC})          & (s)       & Time (s)  & Frames    \\
    \hline
    Ch1     & 2011 Jul 30  & 2455772.6845  & 21421     & 2         & 10496     \\
    Ch2     & 2011 Jul 29  & 2455771.8505  & 21421     & 2         & 10496     \\
   \hline  
\end{tabular}\hfill\strut\ataboff
\end{table}

\section{SECONDARY-ECLIPSE ANALYSIS -- METHODOLOGY}
\label{sec:method}

Exoplanet characterization requires high precision, since the planets' inherently weak signals are weaker than the systematics. In addition, {\em Spitzer}'s systematics lack full physical characterizations. We have developed a  modular pipeline, Photometry for Orbits, Eclipses, and Transits (POET), that implements a wide variety of treatments of systematics and uses Bayesian methods to explore the parameter space and information criteria for model choice. The POET pipeline is documented in our previous papers \citep{StevensonEtal2010Natur, CampoEtal2011apjWASP12b, NymeyerEtal2011apjWASP18b, StevensonEtal2012apjHD149026b, BlecicEtal2013apjWASP14b, CubillosEtal2013apjWASP8b}, so we give here just a brief overview of the specific procedures used in this analysis.


\if\submitms y
\clearpage
\fi
\if\submitms y
\clearpage
\fi

The pipeline uses {\em Spitzer}-supplied BCD frames to produce systematics-corrected light curves and parameter and uncertainty estimates, routinely achieving 85\% of the photon S/N limit or better. Initially, POET masks pixels according to {\em Spitzer's} permanent bad pixel masks, and then it additionally flags bad pixels (energetic particle hits, etc.) by grouping sets of 64 frames and performing a two-iteration, 4\math{\sigma} rejection at each pixel location. Image centers with 0.01 pixel accuracy come from testing a variety of centering routines \citep[][Supplementary Information]{StevensonEtal2010Natur}. Subpixel 5\math{\times} interpolated aperture photometry \citep{HarringtonEtal2007natHD149026b} produces the light curves. We omit frames with bad pixels in the photometry aperture. The background, subtracted before photometry, is an average of good pixels within an annulus centered on the star in each frame. 

Detector systematics vary by channel and can have both temporal (detector ramp) and spatial (intrapixel variability) components. At 3.6 and 4.5 {\micron}, intrapixel sensitivity variation is the dominant effect \citep{CharbonneauEtal2005apjTrES1}, so accurate centering at the 0.01 pixel level is critical. We fit this systematic with a Bilinearly Interpolated Subpixel Sensitivity (BLISS) mapping technique, following \citet{StevensonEtal2012apjHD149026b}, including the method to optimize the bin sizes and the minimum number of data points per bin.

At 8.0, and 16 {\micron}, there is temporal variability, attributed to charge trapping \citep{KnutsonEatl2009ApJ-PhaseVariationHD149026b}. Weak temporal dependencies can also occur at 3.6 and 4.5 {\micron} \citep{Reach2005-IRACCalibration, CharbonneauEtal2005apjTrES1, CampoEtal2011apjWASP12b, DemoryEtal2011apj55Cnc, BlecicEtal2013apjWASP14b}, while weak spatial variability has been seen at 5.8 and 8.0 {\micron} \citep{StevensonEtal2012apjHD149026b, Anderson2011-Ch24-WASP17b}. Thus, we consider both systematics in all channels when determining the best-fit model.

We fit the model components simultaneously using a \citet{MandelAgol2002ApJtransits} eclipse, \math{E(t)}; the time-dependent detector ramp model, \math{R(t)}; and the BLISS map, \math{M(x, y)}:

\begin{eqnarray}
\vspace{-10pt}
\label{eqn:full}
F(x, y, t) = F\sb{s}\,R(t)\,M(x,y)\,E(t),
\end{eqnarray}

\noindent where \math{F(x,y,t)} is the aperture photometry flux and \math{F\sb{\rm s}} is the constant system flux outside of the eclipse.

To choose the best systematics models, we analyze dozens of model combinations and use goodness-of-fit criteria \citep{CampoEtal2011apjWASP12b}. For a given channel, we first vary the photometric aperture size and the number of initial data points that we exclude because of instrument settling, and then we test different ramp models and bin sizes for the intrapixel model.  To choose the best aperture size and the number of initial points dropped during instrument settling, we minimize the standard deviation of the normalized residuals (SDNR). Ignoring data points from the beginning of the observation is a common procedure \citep{Knutson2011-GJ436b} when searching for the best-fitting ramp. We remove the smallest number of points consistent with the minimal SDNR (see each channel analysis for the number of points discarded).

Once we have found the best dataset in this way, we compare different ramp models by applying the Bayesian Information Criterion:

\begin{equation}
{\rm BIC} = \chi\sp{2} + k\ln N,
\label{BIC}
\end{equation}

\noindent where \math{N} is the number of data points. The best model minimizes the chosen criterion. The level of correlation in the photometric residuals is also considered by plotting root-mean-squared (rms) model residuals versus bin size \citep[time interval,][]{pont:2006, winn:2008, CampoEtal2011apjWASP12b} and comparing this to the theoretical \math{1/\sqrt{2N}} rms scaling (\citealp{BlecicEtal2013apjWASP14b} explains the factor of 2). Sometimes, we prefer less-correlated models with insignificantly poorer BIC values.

\begin{figure}[ht]
\vspace{-5pt}
    \centering
    \includegraphics[width=0.99\linewidth, clip]{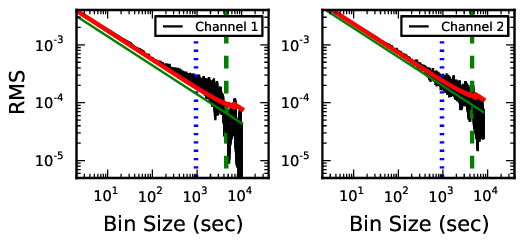}
\vspace{-10pt}
\caption{Correlations of the residuals for the two secondary eclipses of WASP-43b, following \citet{pont:2006}. The black line represents the rms residual flux vs. bin size. The red line shows the predicted standard error scaling for Gaussian noise. The green line shows the theoretical photon noise limit (observed S/N is 80.3\% and 85\% of the photon-limited S/N for channel 1 and channel 2, respectively; see Section \ref{sec:joint}). The black vertical lines at each bin size depict 1\math{\sigma} uncertainties on the rms residuals (\math{rms/\sqrt{2N}}, where N is the number of bins). The dotted vertical blue line indicates the ingress/egress timescale, and the dashed vertical green line indicates the eclipse duration timescale. Large excesses of several \math{\sigma} above the red line would indicate correlated noise at that bin size. Inclusion of 1\math{\sigma} uncertainties shows no noise correlation on the timescales between the ingress/egress duration and the eclipse duration. }
\label{fig:rms}
\end{figure}

We explore the phase space and estimate errors by using a Markov-chain Monte Carlo (MCMC) routine following the Metropolis--Hastings random-walk algorithm, which uses independent Gaussian proposal distributions for each parameter with widths chosen to give an acceptance rate of 30\%--60\%. Each MCMC model fit begins with the Levenberg--Marquardt algorithm (least-squares minimization). We use an informative prior \citep[e.g.,][]{Gelman2002} taken from other work on parameters that are more tightly constrained than what  our fits can achieve. In this work, those are ingress and egress times, which are not well sampled by our observation. All other parameters have flat priors and are free parameters of the MCMC. For each channel, they are listed in Section \ref{sec:joint}. For orbital analysis, they are listed in Section \ref{sec:orbit}. We then run enough MCMC iterations to satisfy the \citet{Gelman1992} convergence test. After every run, we assess convergence by examining plots of the parameter traces, pairwise correlations, autocorrelations, marginal posteriors, best-fitting model, and systematics-corrected best-fitting model. The final fit is obtained from the simultaneous run of all datasets, sharing parameters such as the eclipse midpoint and duration among some or all datasets.

We report the times of our secondary eclipses in both BJD\sb\rm{UTC} (Coordinated Universal Time) and BJD\sb\rm{TT} (BJD\sb\rm{TDB}, Barycentric Dynamical Time), calculated using the Jet Propulsion Laboratory (JPL) Horizons system and following \citet{EastmanEtal2010apjLeapSec}.

\section{SECONDARY-ECLIPSE ANALYSIS -- FIT DETAILS}
\label{sec:results}

Light curves for both channels were extracted using every aperture radius from 2.00 to 4.50 pixels, in 0.25 pixel increments.  We tested three centering routines, center of light, two-dimensional (2D) Gaussian fit, and least asymmetry (see Supplementary Information of \citealp{StevensonEtal2010Natur} and \citealp{LustEtal2013apjCentering}). A 2D Gaussian fit found the most consistent stellar centers. We estimated the background flux by using an annulus of 7--15 pixels from the center of the star for both channels. For the secondary eclipse ingress and egress time, we used a Bayesian prior (\math{t\sb{\rm{2-1}}} = 950.5 {\pm} 145.5 s), calculated from unpublished WASP photometric and RV data. 

Figure \ref{fig:RawBinNorm} shows our systematics-corrected, best-fit light curve models.  Figure \ref{fig:rms} presents the scaling of the rms model residuals versus bin size for both channels, which shows no significant time correlation in the residuals.

\subsection{Channel 1 -- 3.6 {\micron}}
\label{sec:ch1}

The most prominent systematic in this {\em Spitzer}\/ channel is the intrapixel effect. The best BLISS-map bin size is 0.006 pixels when we exclude bins with less than four measurements. The ramp and eclipse models fit without removing initial data points. The smallest value of BIC reveals that the best ramp model is quadratic; this model is 1.2\tttt{30} times more probable than the next-best (linear) model. Table \ref{table:Ch1-ramps} lists the best ramp models, comparing their SDNR, BIC values, and eclipse depths.


\begin{table}[ht]
\caption{\label{table:Ch1-ramps} 
Channel 1 Ramp Models}
\atabon\strut\hfill\begin{tabular}{lcrcc}
    \hline
    \hline
    Ramp Model    & SDNR           & \math{\Delta}BIC  & Eclipse Depth (\%) \\
    \hline
    Quadratic    & 0.0039001       & 0.0        &  0.344 {\pm} 0.013 \\
    Rising       & 0.0039113       & 56.8       &  0.292 {\pm} 0.012 \\
    No-Ramp      & 0.0039315       & 144.4      &  0.268 {\pm} 0.012 \\
    Linear       & 0.0039293       & 142.1      &  0.270 {\pm} 0.012 \\
    \hline
\end{tabular}\hfill\strut\ataboff
\end{table}

\begin{figure}[ht]
    \centering
    \vspace{-10pt}
    \includegraphics[width=0.9\linewidth, clip]{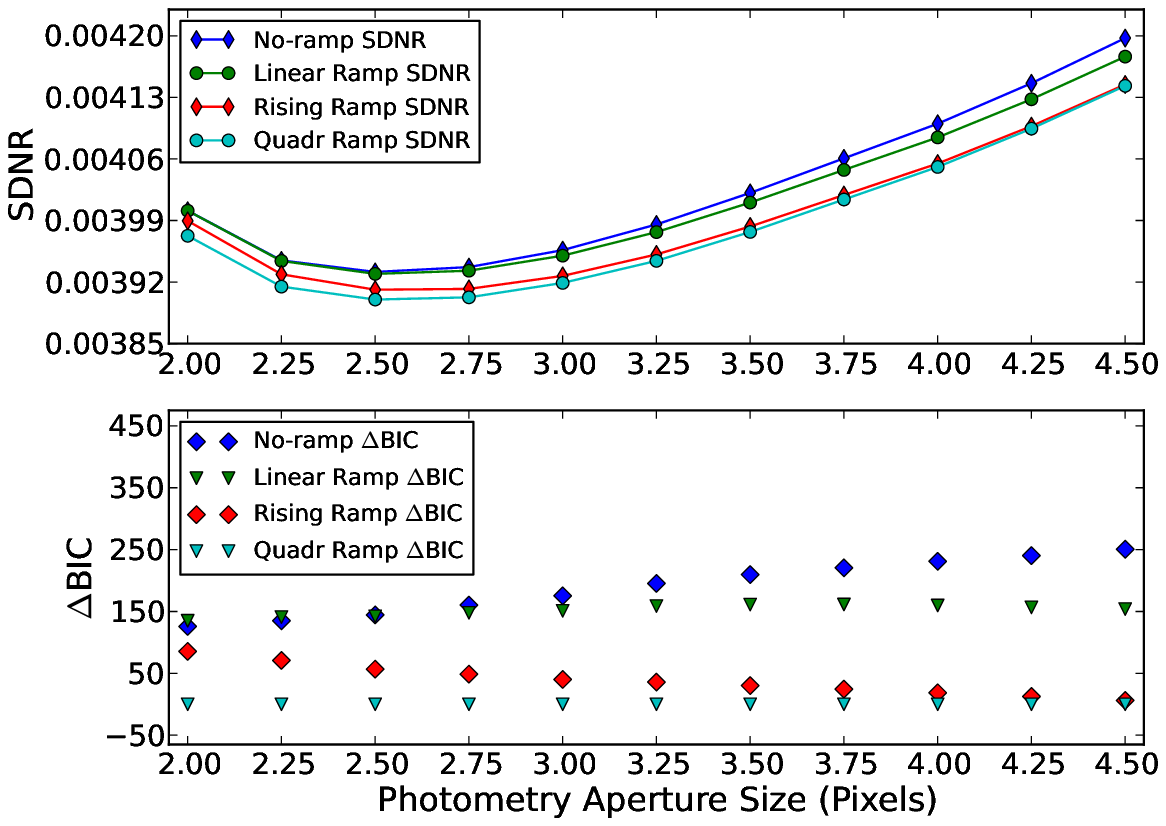}\vspace{-3pt}
\caption{
Channel 1 comparison between different ramp models. The plots show SDNR vs. aperture size and \math{\Delta}BIC vs. aperture size.  A lower SDNR value indicates a better model fit. The lowest SDNR value marks the best aperture size (2.50 pixels). A lower \math{\Delta}BIC value at the best aperture size indicates which ramp model is the best (quadratic ramp model, green triangles).}
\label{fig:ch1-SDNR}
\end{figure}

Figure \ref{fig:ch1-SDNR} shows a comparison between the best ramp models and their SDNR and BIC values through all aperture sizes, indicating which aperture size is the best and which model has the lowest BIC value. 

\begin{figure}[ht]
    \vspace{-5pt}
    \centering
    \includegraphics[width=0.9\linewidth, clip]{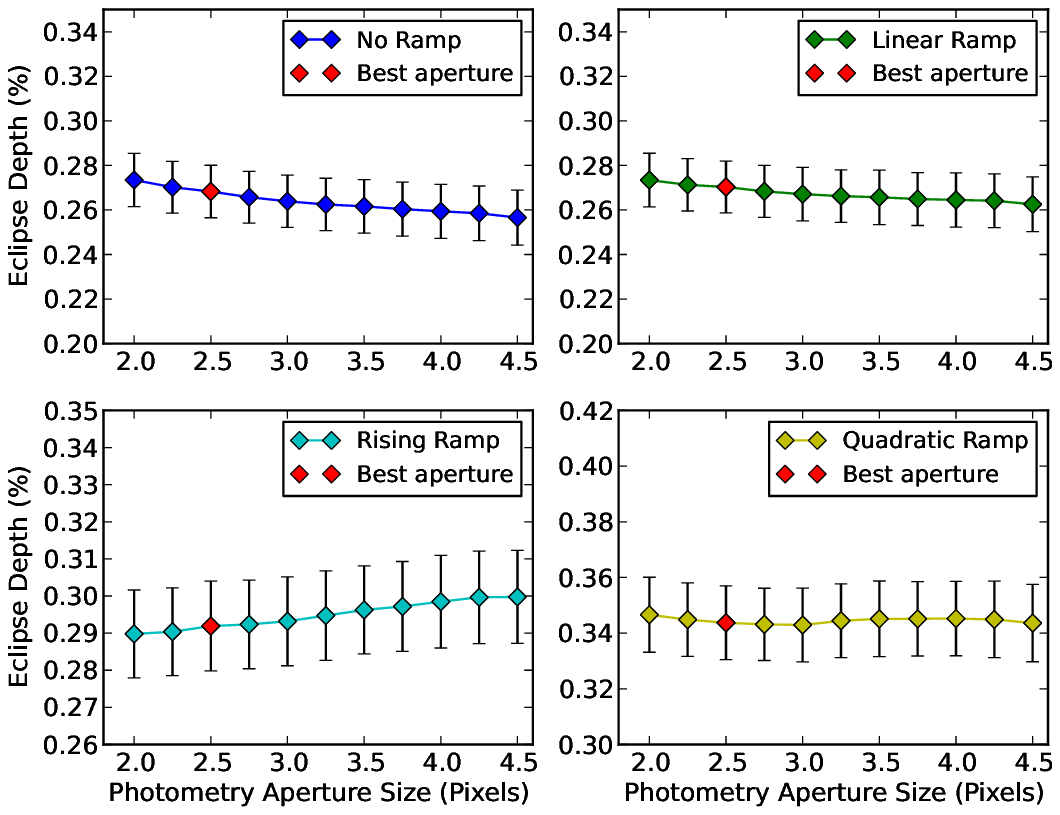}
\caption{
Best-fit eclipse depths as a function of photometry aperture size for channel 1. The four best ramp models are plotted. The red point indicates the best aperture size for that channel. The eclipse depth uncertainties are the result of 10\sp{5} MCMC iterations. The trend shows insignificant dependence of eclipse depth on aperture size (much less than 1\math{\sigma}).}
\label{fig:ch1-depths}
\end{figure}

Photometry generates consistent eclipse depths for all tested apertures, with the lowest SDNR at an aperture radius of 2.50 pixels (Figure \ref{fig:ch1-depths}). 

\subsection{Channel 2 -- 4.5 {\micron}}
\label{sec:ch2}

\begin{figure}[ht]
    \centering
    \vspace{-10pt}
    \includegraphics[width=0.9\linewidth, clip]{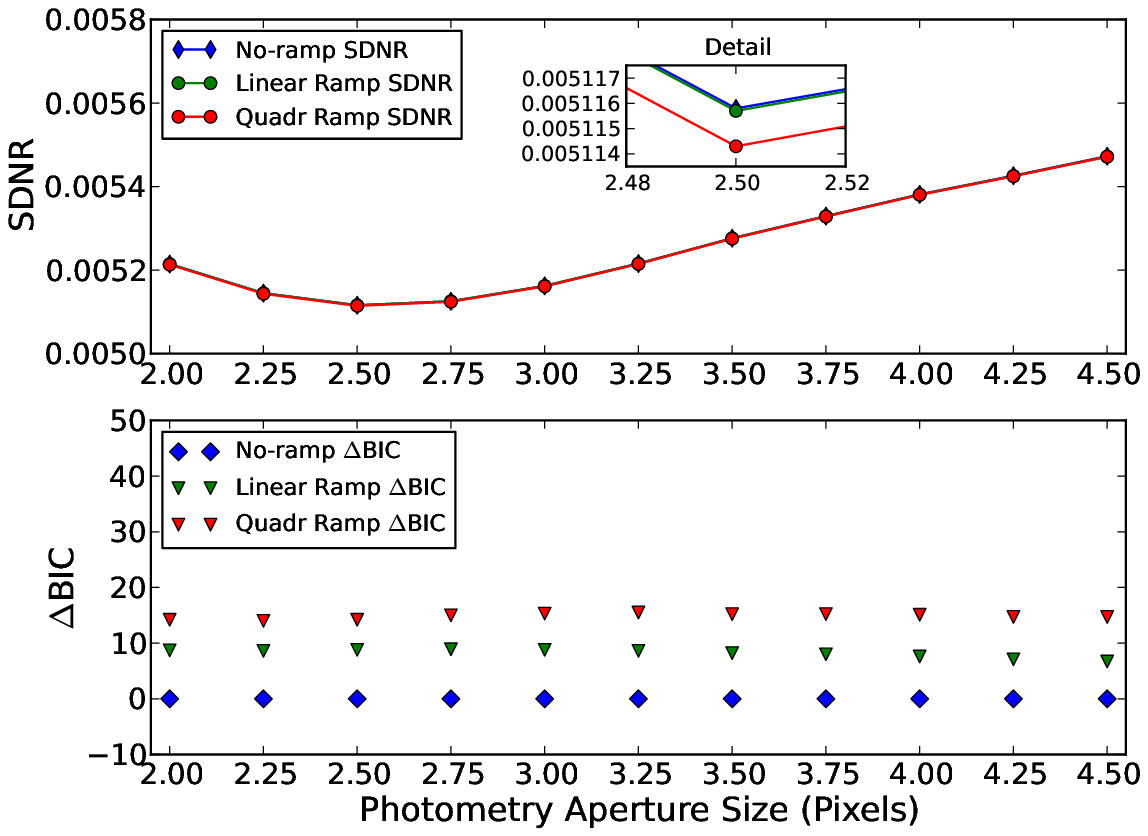}\vspace{-3pt}
\caption{ Channel 2 comparison between different ramp models. Plot shows SDNR vs. aperture size and \math{\Delta}BIC vs. aperture size.  A lower SDNR value indicates a better model fit. The lowest SDNR value marks the best aperture size (2.50 pixels). Lower BIC values at the best aperture size indicate better models (best: no-ramp model, blue diamonds). The inset shows separation in SDNR for different ramp models at the best aperture size. 
}

\label{fig:ch2-SDNR}
\end{figure}

In this channel, we noticed an upward trend in flux at the beginning of the observation, possibly due to telescope settling, which we do not model.  We clipped 2300 initial data points (\sim 38 minutes of observation), the smallest number of points consistent with the minimal SDNR. The 2.50 pixel aperture radius minimizes SDNR (Figure \ref{fig:ch2-SDNR}). 

To remove intrapixel variability, the BLISS bin size is 0.016 pixels, ignoring bins with less than four points. The lowest BIC value corresponds to the model without a ramp (Table \ref{table:Ch2-ramps}), which is 78 times more probable than the linear model. 

\begin{table}[ht]
\caption{\label{table:Ch2-ramps} 
Channel 2 Ramp Models}
\atabon\strut\hfill\begin{tabular}{lcccc}
    \hline
    \hline
    Ramp Model   & SDNR           & \math{\Delta}BIC    & Eclipse Depth (\%)      \\
    \hline
    No-Ramp      & 0.0051158      &  0                   & 0.392 {\pm} 0.016    \\
    Linear       & 0.0051157      &  8.8                 & 0.392 {\pm} 0.016    \\
    Quadratic    & 0.0051143      &  14.2                & 0.409 {\pm} 0.019    \\ 
    \hline
\end{tabular}\hfill\strut\ataboff
\end{table}

\begin{figure}[h]
    \vspace{-5pt}
    \centering
    \includegraphics[width=0.8\linewidth, clip]{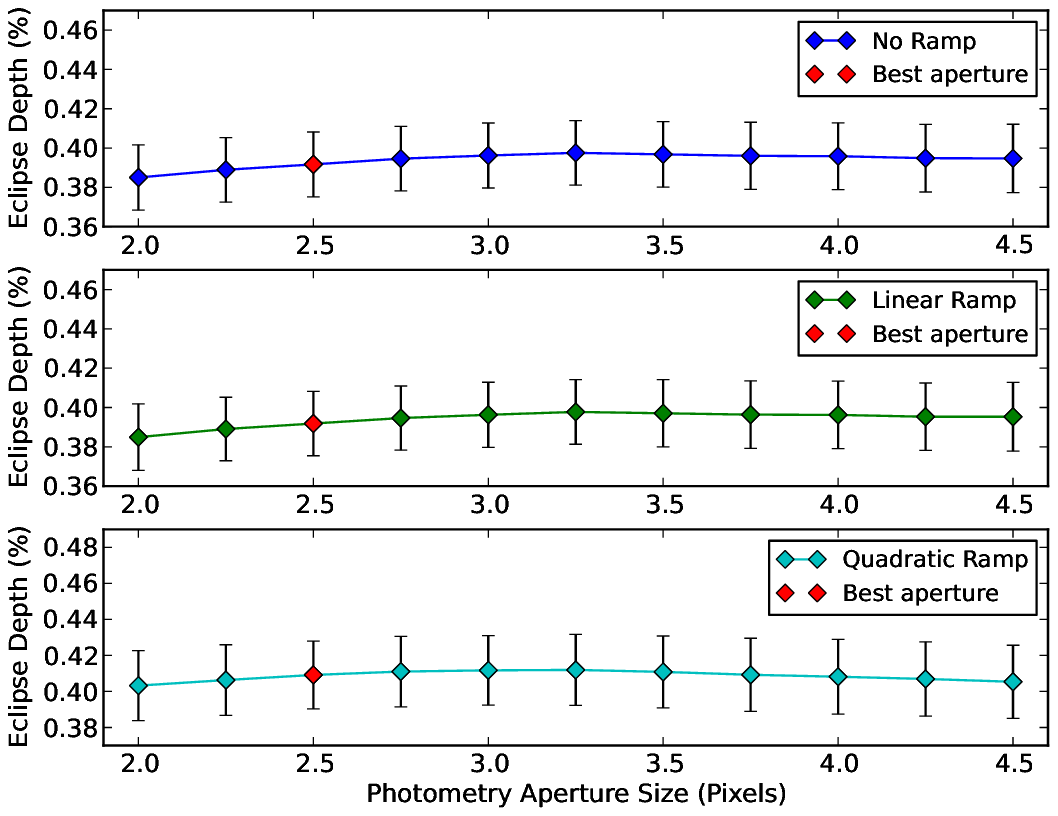}
\caption{
Best-fit eclipse depths as a function of photometry aperture size for channel 2. The three best ramp models are plotted. The red point indicates the best aperture size for that channel. The eclipse-depth uncertainties are the result of 10\sp{5} MCMC iterations. The trend shows negligible dependence of eclipse depth on aperture size (much less than 1\math{\sigma}).}
\label{fig:ch2-depths}
\end{figure}

We tested the dependence of eclipse depth on aperture radius, showing that they are all well within 1\math{\sigma} of each other, to validate the consistency of our models (Figure \ref{fig:ch2-depths}).

\subsection{Joint Fit}
\label{sec:joint}

To improve accuracy, we share the eclipse width (duration), eclipse midpoint phases, and ingress and egress times in a joint fit of both datasets. Table 4 indicates which parameters are free, shared, or have informative priors. The best ramp models and the best aperture sizes from the separate channel analyses are used in the joint fit. To produce the best joint-model fits, we iterated MCMC until the \citet{Gelman1992} diagnostics for all parameters dropped below 1\%, which happened after \ttt{5} iterations. 

The best joint-model fit parameters are in Table \ref{tab:eclfits}.  Files containing the light curves, best model fits, centering data, photometry, etc., are included as electronic supplements to this article.  The eclipse midpoint time is further used for the subsequent orbital analysis, and the eclipse depths are used for the atmospheric analysis.

\begin{table*}[ht]
\centering
\caption{\label{tab:eclfits} Best-fit Joint Eclipse Light curve Parameters}
\begin{tabular}{lr@{\,{\pm}\,}lr@{\,{\pm}\,}lr@{\,{\pm}\,}l}
\hline
\hline
\colhead{Parameter}                                                  &   \mctc{    Channel 1     }           &   \mctc{   Channel 2      }     \\
\hline
Array position (\math{\bar{x}}, pixel)                                 &   \mctc{      14.99       }           &   \mctc{       14.8       }     \\
Array position (\math{\bar{y}}, pixel)                                 &   \mctc{      15.01       }           &   \mctc{      15.05       }     \\
Position consistency\tablenotemark{a} (\math{\delta\sb{x}}, pixel)     &   \mctc{      0.009       }           &   \mctc{      0.016       }     \\
Position consistency\tablenotemark{a} (\math{\delta\sb{y}}, pixel)     &   \mctc{      0.012       }           &   \mctc{      0.014       }     \\
Aperture size (pixel)                                                  &   \mctc{      2.5         }           &   \mctc{      2.5         }     \\
Sky annulus inner radius (pixel)                                       &   \mctc{      7.0         }           &   \mctc{      7.0         }     \\
Sky annulus outer radius (pixel)                                       &   \mctc{      15.0        }           &   \mctc{      15.0        }     \\
System flux \math{F\sb{s}} (\micro Jy)\tablenotemark{b}              &         64399.0 & 5.0                 &         37911.0 & 2.0           \\
Eclipse depth (\%)\tablenotemark{b}                                  &           0.347 & 0.013               &         0.382   & 0.015         \\
Brightness temperature (K)                                           &           1670 & 23                &             1514   & 25          \\
Eclipse midpoint (orbits)\tablenotemark{b}                           &          0.4986 & 0.0004              &          0.4986 & 0.0004        \\
Eclipse midpoint (BJD\sb\rm{UTC}--2,450,000)                           &       5773.3172 & 0.0003              &       5772.5037 & 0.0003        \\
Eclipse midpoint (BJD\sb\rm{TDB}--2,450,000)                           &       5773.3179 & 0.0003              &       5772.5045 & 0.0003        \\
Eclipse duration (\math{t\sb{\rm{4-1}}}, hr)\tablenotemark{b}       &            1.25 & 0.02                &            1.25 & 0.02          \\
Ingress/egress time (\math{t\sb{\rm{2-1}}}, hr)\tablenotemark{b}    &           0.268 & 0.018               &           0.268 & 0.018         \\
Ramp name                                                            &   \mctc{        quadramp  }           &   \mctc{     no-ramp      }     \\
Ramp, quadratic Term\tablenotemark{b}                                &         -0.0827 & 0.0069              &   \mctc{      ...         }      \\
Ramp, linear term\tablenotemark{b}                                   &         -0.0002 & 0.0005              &   \mctc{      ...         }      \\
Intrapixel method                                                    &   \mctc{       BLISS      }           &   \mctc{      BLISS       }     \\
BLISS bin size in $x$  (pixel)                                         &   \mctc{     0.006        }           &   \mctc{     0.016        }     \\
BLISS bin size in $y$  (pixel)                                         &   \mctc{     0.006        }           &   \mctc{     0.016        }     \\
Minimum number of points per bin                                     &   \mctc{        4         }           &   \mctc{        4         }     \\
Total frames                                                         &   \mctc{      10496       }           &   \mctc{      10496       }     \\
Frames used                                                          &   \mctc{      10124       }           &   \mctc{       8004       }     \\
Rejected frames (\%)                                                 &   \mctc{  0.44            }           &   \mctc{  1.12            }     \\
Free parameters                                                      &   \mctc{        7         }           &   \mctc{        2         }     \\
Number of data points in fit                                         &   \mctc{      10124       }           &   \mctc{       8004       }     \\
BIC                                                                  &   \mctc{  18207.3         }           &   \mctc{  18207.3         }     \\
SDNR                                                                 &   \mctc{ 0.0039007        }           &   \mctc{ 0.0051167        }     \\
Photon-limited S/N (\%)                                              &   \mctc{       80.3       }           &   \mctc{       85.0       }     \\
\hline\end{tabular}
\begin{minipage}[t]{0.65\linewidth}
\tablenotetext{1}{rms frame-to-frame position difference.}
\tablenotetext{2}{Free parameter in MCMC fit.  All priors are flat except
\math{t\sb{2-1}}, which uses a Gaussian prior of 950.5 {\pm} 145.5 s, calculated from unpublished WASP photometric and RV data. Parameters
with identical values and uncertainties are fit jointly.}
\end{minipage}
\end{table*}
\if\submitms y
\clearpage
\fi

\section{ORBIT}
\label{sec:orbit}

The eclipse midpoint (after a 15.2 s correction for the eclipse transit light-time) has a phase of 0.5001 {\pm} 0.0004, so \math{e \cos \omega} = 0.0001 {\pm} 0.0006, or a \math{3\sigma} upper limit of \math{|e \cos \omega| < 0.0018}, consistent with a circular orbit. 


\begin{figure}[hb]
    \vspace{-5pt}
    \centering
    \includegraphics[width=0.9\linewidth, clip]{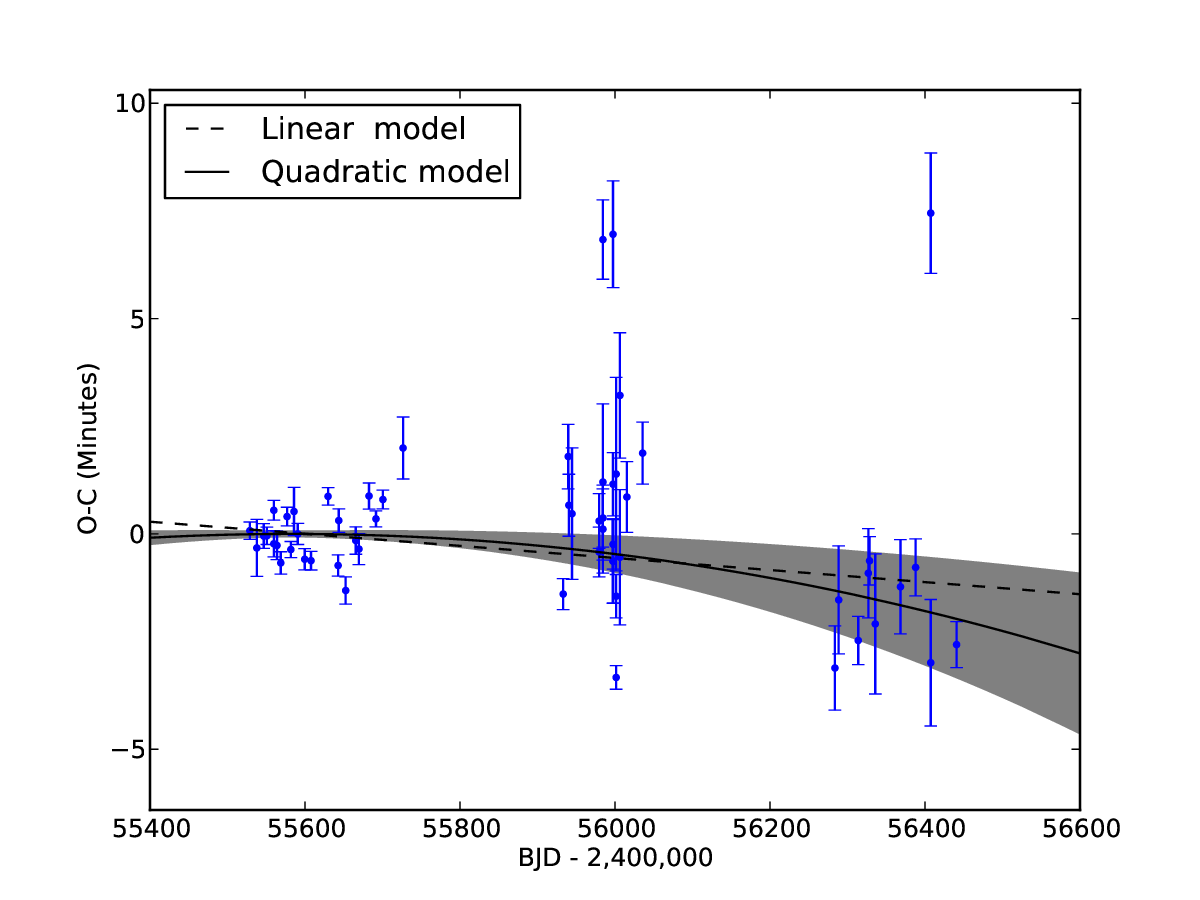}
\caption{O-C diagram for transit observations of WASP-43b with respect to the linear terms in the best-fit quadratic ephemeris.  The quadratic ephemeris is shown with the 1\math{\sigma} prediction uncertainty in grey.  The best-fit linear ephemeris is also shown as a dashed line.    Only points with an Exoplanet Transit Database quality rating of 3 or better are shown here and used in this analysis. }
\label{fig:ttv}
\end{figure}

To improve the orbit solution further, we combined data from our observations with data from a variety of sources (see Table \ref{tab:ttv}).  Transit midpoint times were taken from \citet{Hellier2011-WASP43b} and \citet{GillonEtal2012A&A-WASP-43b}, and amateur observations were listed in the Exoplanet Transit Database (see Table \ref{tab:ttv}).  We used {\em CORALIE} RV observations published by \citet{Hellier2011-WASP43b} and \citet{GillonEtal2012A&A-WASP-43b}.  No RV points analyzed were gathered during transit. We subtracted 15.2 s from the eclipse midpoint to correct for light-travel time across the orbit. We corrected all points to TDB if this was not already done \citep{EastmanEtal2010apjLeapSec}.  We converted the amateur data from HJD to BJD, putting all times in a consistent BJD\sb\rm{TDB} format.  There were 49 transit points, 23 RV points, and one effective eclipse observation.  We fit all of these data simultaneously, as described by \citet{CampoEtal2011apjWASP12b}.  The free parameters in this fit were \math{e \sin \omega}; \math{e \cos \omega}; the period, \math{P}; the reference transit midpoint time, \math{T\sb{\rm{0}}}; the RV semi-amplitude, \math{K}; and the RV offset, \math{\gamma}.  The addition of the amateur transit observations improves the uncertainty of \math{P} by a factor of nearly five compared with \citet{GillonEtal2012A&A-WASP-43b}, reducing it to 13 ms.  The fit finds an eccentricity of 0.010$^{+0.010}_{-0.007}$, consistent with a circular orbit and expectations for a close-in planet, where eccentricity should be damped by tidal interactions with the host star \citep{JacksonEtal2008ApJ-TidalEvolution}. Table \ref{tab:orbit} summarizes the fit results.

Because \math{e \sin \omega} is a much larger component of the eccentricity than \math{e \cos \omega},  it is possible that much of this eccentricity signal comes from the effect of the planet raising a tidal bulge on its host star. \citet{ArrasEtal2012MNRAS-RVstar} predict that the RV semi-amplitude of this effect is 8.9 ms\sp{-1}. Since our model shows that \math{eK} = 5$^{+6}_{-3}$ ms\sp{-1}, it is possible that the majority or entirety of the eccentricity signal is due to the tidal bulge interaction, and the true eccentricity is closer to the upper limit derived from the secondary eclipse.

We found that, for a linear ephemeris fit to just the transit timing data, there is considerable scatter in O-C (observed time minus calculated time).  The root-mean-square of the stated transit-time uncertainties is 51 s, while the standard deviation of the residuals is 124 s.  WASP-43b is close enough to its host star that tidal decay is a significant factor in its evolution, so we attempted to estimate the decay rate by adding a quadratic term to our ephemeris model, following \citet{AdamsEtal2010ApJ}.  Our model for the transit ephemeris is now:

\begin{equation}
T\sb{N} = T\sb{0} + PN + \delta P \frac{N(N-1) }{2},
\end{equation}

\noindent where \math{N} is the number of orbits elapsed since the epoch \math{T\sb{\rm{0}}}, \math{P} is the orbital period at \math{T\sb{\rm{0}}}, and \math{\delta P = \dot{P}P}, where \math{\dot{P}} is the short-term rate of change in the orbital period.  Fitting this model to the transit data, we find that \math{\delta P} = (-2.5 {\pm} 0.9)\tttt{-9} days orbit\sp{-2}, or \math{\dot{P}} = -0.095 {\pm} 0.036 s yr\sp{-1}.  This is illustrated in Figure \ref{fig:ttv}.   This is a nondetection, though the best-fit value is comparable to the value of -0.060 {\pm} 0.015 s yr\sp{-1} found by  \citet{AdamsEtal2010ApJ} for OGLE-TR-113b.

While the suggestion of a quadratic inspiral existed in the data
gathered prior to BJD 2456035, the addition of 17 new amateur
observations from BJD 2456250 to BJD 2456440 (11 of which were of
sufficiently high data quality for inclusion) put the linear fit within
the credible region of the quadratic fit.  

Table \ref{tab:ephem} summarizes all three transit ephemeris models. The first is the linear ephemeris from the fit discussed above. We also fit linear and quadratic models to just the transit data to study any trend in orbital period. These are the second and third fits in Table \ref{tab:ephem}. Their BIC and chi \math{\chi\sp{2}} are comparable to each other but not to the first fit, since the data sets are not identical.

\begin{table}[ht]
\centering
\caption{\label{tab:ttv} Transit Timing Data}
\begin{tabular}{lll@{ }c@{ }}
\hline
\hline
Mid-transit Time  &  Uncertainty          &   Source\tablenotemark{a}  & Quality  \\
 (BJD\sb\rm{TDB})    &                       &                            & Rating   \\
\hline
2456440.77250		& 0.00037		& Phil Evans							 & 1 	\\
2456410.67704	   	& 0.00097		& Robert Majewski				 		 & 5	\\
2456407.42697		& 0.00097		& Enrique D\'{i}ez Alonso			 	         & 3	\\
2456407.41972		& 0.00102		& Ullrich Dittler						 & 3	\\
2456403.34716		& 0.0017		& Jens Jacobsen							 & 5	\\
2456401.72885		& 0.00199		& Alex Chassy							 & 5	\\
2456387.89785		& 0.00046		& Phil Evans							 & 2	\\
2456375.70739		& 0.00158		& Parijat Singh							 & 4	\\
2456368.37413		& 0.00076		& Adam B\"uchner							 & 3	\\	
2456335.83452		& 0.00113		& Phil Evans							 & 3	\\
2456328.51426		& 0.00039		& Juan Lozano de Haro					         & 2	\\
2456328.51296		& 0.00181		& Daniel Staab							 & 4	\\
2456326.88711		& 0.00072		& Phil Evans							 & 3	\\
2456313.87042		& 0.00039		& P. Kehusmaa \& C. Harlingten	                                 & 2	\\
2456288.65334		& 0.00087		& Jordi Lopesino						 & 3	\\
2456283.77139		& 0.00068	        & A. Chapman \& N. D. D\'{i}az		                         & 3	\\
2456250.42005		& 0.00067	        &L. Zhang, Q. Pi \& A. Zhou                                      & 5	\\
2456035.66489       & 0.0005        & George Hall                         & 2         \\
2456015.3273        & 0.00057       & Martin Z\'{i}bar                    & 2 	     \\
2456006.38071       & 0.00101       & Franti\^{s}ek Lomoz                 & 3	     \\
2456006.3781        & 0.00109       & Franti\^{s}ek Lomoz                 & 3	     \\
2456001.49859       & 0.00156	    & Alfonso Carre\~{n}o                     & 3	     \\
2456001.49662       & 0.00035       & Gustavo Muler Schteinman            & 1	     \\
2456001.49531       & 0.00019       & Fernand Emering                     & 1	     \\
2455997.43508       & 0.00086       & Ren\'{e} Roy                        & 3	     \\
2455997.43105       & 0.00051       & Faustino Garcia                     & 3         \\
2455997.43008       & 0.0004        & Nicolas Esseiva             	 & 2	     \\
2455997.42981       & 0.00068       & Juanjo Gonzalez             	 & 3	     \\
2455984.41939       & 0.00064       & Ferran Grau Horta           	 & 3	     \\
2455984.41548       & 0.00126       & Franti\^{s}ek Lomoz            & 3	     \\
2455984.4149        & 0.00047       & Fabio Martinelli            	 & 2         \\
2455984.41472       & 0.00071       & Franti\^{s}ek Lomoz          	 & 3	     \\
2455979.534         & 0.00044       & Nicolas Esseiva            	 & 2	     \\
2455979.5335        & 0.0004        & Juanjo Gonzalez            	 & 2	     \\
2455957.57296       & 0.00122       & Franti\^{s}ek Lomoz          	 & 4	     \\
2455944.55468       & 0.00106       &  Roy Ren\'{e}            		 & 3	     \\
2455940.48744       & 0.0005        & Anthony Ayiomamitis         	 & 2	     \\
2455939.67475       & 0.00052       & Ramon Naves             		 & 2	     \\
2455933.16473       & 0.00025       & Peter Starr            		 & 1	     \\
2455686.68399       & 0.0008        & Stan Shadick           		 & 3	     \\
2455682.61364       & 0.00039       & Tanya Dax, Stacy Irwin       	 & 5	     \\
\hline
\end{tabular}
\begin{minipage}[t]{0.95\linewidth}
\tablenotetext{1}{The TRansiting ExoplanetS and CAndidates group (TRESCA, http://var2.astro.cz/EN/tresca/index.php) supply their data to the Exoplanet Transit Database (ETD, http://var2.astro.cz/ETD/), which performs the uniform transit analysis described by \citet{Poddany2010}.  The ETD web site provided the numbers in this table, which were converted from HJD (UTC) to BJD (TDB).}
\end{minipage}
\end{table}
\if\submitms y
\clearpage
\fi

The BIC values still favor the quadratic
(decaying) ephemeris (BIC = 450) over the linear ephemeris (BIC = 452.8)
by a probability ratio of \math{e}\sp{\math{\Delta}BIC/2} = 4.  Two lines of reasoning
favor the linear ephemeris, however, so we consider the linear ephemeris
to be more likely.

First, the inspiral time predicted by the quadratic is extremely short
compared to the planet's lifetime.  We would need to believe that we are
seeing the planet at a very brief and special time in its history to
accept the conclusion of inspiral.

Second, the rms scatter about the quadratic ephemeris is 121 s, much
larger than the 51 s typical transit time uncertainty and not much
different from the linear result. The reduced \math{\chi\sp{2}} for the quadratic fit
is \sim 8, so much of the residual scatter is unexplained by either
model. Possible explanations include stellar activity, transit timing
variations (TTVs), or problems in data processing or reporting. Mistakes
in the time corrections for heterogeneous transit data are unlikely
because the Exoplanet Transit Database indicates that all amateur
observations were submitted in UTC, while the professional data were
unambiguous in their use of TDB.  While it is possible that
uncertainties for certain sets of transit data points may have been
underestimated, the data come from many amateur and professional
sources, and all would have had to make such errors. Since WASP-43 is an
early K-type star, a likely explanation for the scatter is stellar
activity.

Effects like TTVs or tidal infall could still contribute to the scatter,
so we present related calculations below, mainly as motivation and
background for future studies.

\begin{table}[ht]
\centering
\caption{\label{tab:orbit} Eccentric Orbital Model}
\begin{tabular}{lc}
\hline
\hline 
Parameter                                                          &  Value                                  \\
\hline
\math{e \sin \omega}\tablenotemark{a}     	                   &  \math{-0.010}    {\pm} 0.011           \\
\math{e \cos \omega}\tablenotemark{a}     	                   &  \math{-0.0003}   {\pm} 0.0006         \\
\math{e}                          			           &  0.010           $^{+0.010}_{-0.007}$  \\
\math{\omega} (\degree)           			           &  \math{-88}       $^{+5}_{-9}$          \\
\math{P} (days)\tablenotemark{a}      		                   &  \math{P} = 0.81347436 {\pm} 1.4\math{\times 10\sp{-7}} \\
\math{T\sb{\rm 0}} (BJD\sb\rm{TDB}) \tablenotemark{a}                 &  2455528.86857    {\pm} 0.00005         \\
\math{K} (m\,s\sp{-1})\tablenotemark{a}                            &  549              {\pm} 6               \\
\math{\gamma} (m\,s\sp{-1})\tablenotemark{a}                       &  \math{-3595}     {\pm} 4               \\
\math{\chi^2}				                           &  458                                     \\
\hline                                           
\end{tabular}
\begin{minipage}[t]{0.55\linewidth}
\tablenotemark{a}{Free parameter in MCMC fit.}\comment{\\
\tablenotemark{b}{BJD\sb{TDB}.}\\
\tablenotemark{c}{RV semi-amplitude}.\\
\tablenotemark{d}{RV offset}}
\end{minipage}
\end{table}
\if\submitms y
\clearpage
\fi

\begin{table*}[ht]
\centering
\caption{\label{tab:ephem} Ephemeris Solutions}
\begin{tabular}{lccc}
\hline
\hline 
                                        & All Data (Transits, Eclipses, RV)	& Transits Only   & Transits Only\\
Parameter                       &  Linear Ephemeris                         & Linear Ephemeris  & Quadratic Ephemeris              \\
\hline
\math{T_0} (BJD\sb\rm{TDB})    	&  2455528.86857    {\pm} 0.00005  &  2455581.74439    {\pm} 0.00004   & 2455581.74437 {\pm} 0.00004    \\
\math{P} (days orbit\sp{-1})    &  0.81347436       {\pm} 1.4\math{\times 10\sp{-7}} & 0.81347450  {\pm} \math{1.5\times 10\sp{-7}}    & 0.81347530   {\pm} 3.8\math{\times 10\sp{-7}}\\
\math{\delta P} (days orbit\sp{-2})\tablenotemark{a}   &  ...  &         ...         &     \math{(-2.5 {\pm} 0.9) \times 10\sp{-9}}\\
\math{\dot{P}\,} (s\,\,yr\sp{-1})\tablenotemark{b}    &  ...  &        ...          &      \math{-0.095 {\pm} 0.036 }      \\
\math{\chi^2}			   &               ...                          &        444.7                  &	   437.9	     \\
BIC                 	           &               ...                  &        452.8                   &    450.0              \\
\hline
\multicolumn{3}{l}{\tablenotemark{a}{\math{\delta P = \dot{P}P}}. 
\tablenotemark{b}{Derived parameter}.}
\end{tabular}
\end{table*}
\if\submitms y
\clearpage
\fi

From the measured period change of \math{\dot{P}} = -0.095 {\pm} 0.036  s yr\sp{-1}, we adopt a 3\math{\sigma} upper limit, \math{|\dot{P}|} < 0.129 s yr\sp{-1}. For WASP-43b, this translates to a maximum change in the semimajor axis of \math{|\dot{a}|} < 1.9 \tttt{-8} AU yr\sp{-1}.  The three-sigma upper limit on the period decay also suggests an infall timescale of at least 5\tttt{5} ys.  \citet{LevrardEtAl2009ApJ-FallingHotJupiters} give a relation for tidal decay, which for synchronous planetary rotation and negligible eccentricity and obliquity reduces to:

\begin{equation}
\frac{1}{a}\frac{{\rm d} a}{{\rm d} t} = \frac{6}{Q'\sb\star }\frac{M\sb{p}}{M\sb\star}\left(\frac{R\sb\star}{a}\right)\sp5 \left(\omega\sb\star-\frac{2\pi}{P}\right),
\end{equation}

\noindent where \math{Q'\sb{\rm{\star}}} is the ratio of the stellar tidal quality factor to the second-order stellar tidal Love number, \math{k\sb{\rm{2}}}, and \math{\omega\sb{\rm{\star}}} is the stellar rotation rate. The upper limit on the quadratic term implies \math{Q'\sb\star > 12,000}. This is much lower than the values of \ttt{5}--\ttt{10} normally assumed and thus cannot rule out any plausible values.  A small value of \math{Q'\sb{\rm{\star}}} was also found by \cite{AdamsEtal2010ApJ}.

While the quadratic fit failed to produce a useful upper limit on tidal decay, observations with a longer time baseline may yet find secular changes or TTVs.  Until then, the linear ephemeris presented here is the most reliable predictor of future transit times.

\section{ATMOSPHERE}
\label{sec:atm}

We modeled the dayside atmosphere of WASP-43b by using the atmospheric modeling and retrieval method of \citet{MadhusudhanSeager2009, MadhusudhanSeager2010}. The model computes line-by-line radiative transfer in a one-dimensional, plane-parallel atmosphere, with constraints of local thermodynamic equilibrium, hydrostatic equilibrium, and global energy balance. The pressure--temperature profile and components of the molecular composition are free parameters of the model, allowing exploration of models with and without thermal inversions and those with oxygen-rich as well as carbon-rich compositions \citep{Madhusudhan2012}. 

The model includes all the primary sources of opacity expected in hydrogen-dominated giant-planet atmospheres in the temperature regimes of hot Jupiters, such as WASP-43b. The opacity sources include line-by-line absorption due to H\sb{2}O, CO, CH\sb{4}, CO\sb{2}, and NH\sb{3} and collision-induced absorption (CIA) due to H\sb{2}-H\sb{2}. We also include hydrocarbons besides CH\sb{4}, such as HCN and C\sb{2}H\sb{2}, which may be abundant in carbon-rich atmospheres \citep{Madhusudhan2011b, Kopparapu2012ApJ-CarbonRich, Madhusudhan2012}. Since in highly irradiated oxygen-rich atmospheres TiO and VO may be abundant \citep{Fortney2008}, we also include line-by-line absorption due to TiO and VO in regions of the atmosphere where the temperatures exceed the corresponding condensation temperatures. Our molecular line data are from \citet{Freedman08}, R. S. Freedman (2009, private communication), \citet{Rothman2005-HITRAN}, \citet{KarkoschkaTomasko2010}, E. Karkoschka (2011, private communication), and \citet{HarrisEtal2008MNRAS-CarbonStars}. We obtain the H\sb{2}-H\sb{2} CIA opacities from \citet{Borysow1997} and \citet{Borysow2002}. The volume mixing ratios of all the molecules are free parameters in the model. 

We constrain the thermal structure and composition of WASP-43b by combining our {\em Spitzer} photometric observations at 3.6 {\micron} and 4.5 {\micron} combined with previously reported ground-based narrow-band photometric data from \citet{GillonEtal2012A&A-WASP-43b},  obtained using VLT/HAWK-I at 1.19 {\micron} and 2.09 {\micron}, and broadband photometric data from \citet{WangEtal2013-WASP43b} obtained using CFHT/WIRCAMthe in the \math{H} (1.6 {\micron}) and \math{K\sb{s}} (2.1 {\micron}) bands.  
The data also place a joint constraint on the day--night energy redistribution and the Bond albedo by requiring global energy balance, i.e., that the integrated emergent power from the planet does not exceed the incident irradiation.  Given that the number of model parameters is \math{\geq} 10 (depending on the C/O ratio) and the number of available data points is 6, our goal is to find the regions of model space favored by the data, rather than to determine a unique fit. We explore the model parameter space by using an MCMC routine (for details, see \citealp{MadhusudhanSeager2009, MadhusudhanSeager2010, MadhusudhanEtal2011natWASP12batm}).


\begin{figure*}[ht]
\centerline{
\includegraphics[width = 0.68\textwidth, clip]{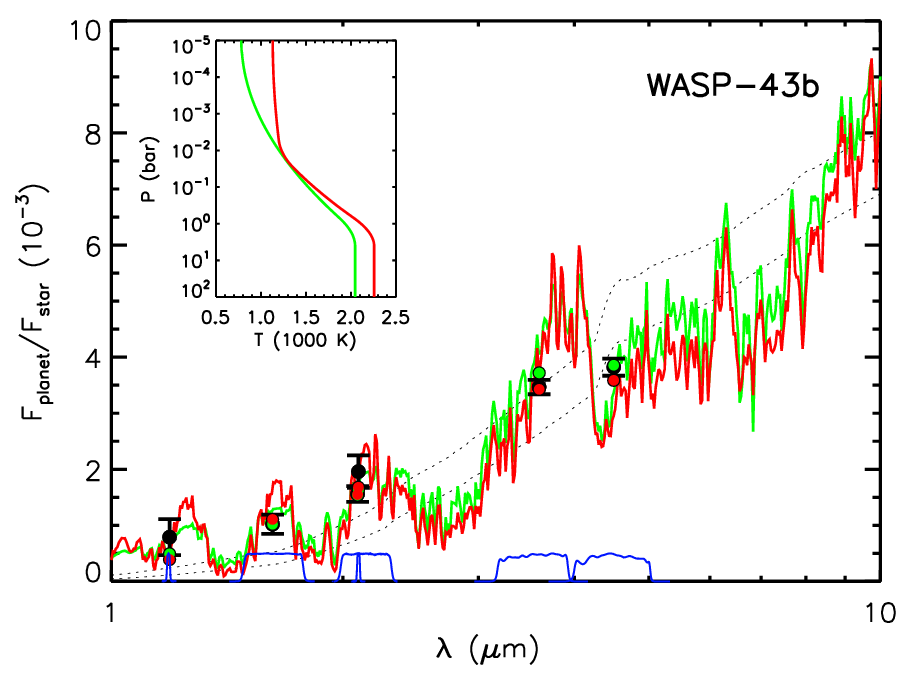}}
\caption{
Observations and model spectra for dayside thermal emission from WASP-43b. The black filled circles with error bars show our data in {\em Spitzer} IRAC channels 1 (3.6 {\micron}) and 2 (4.5 {\micron}) and previously published ground-based near-infrared data in narrow-band photometry at 1.19 {\micron} and 2.09 {\micron} \citep{GillonEtal2012A&A-WASP-43b} and in broadband photometry at 1.6 {\micron} and 2.1 {\micron} \citep{WangEtal2013-WASP43b}. The solid curves show the model spectra in the main panel, and the corresponding temperature--pressure profiles, with no thermal inversions, in the inset. The green and red curves correspond to models with compositions of nearly solar and 10 \math{\times} solar metallicity, respectively.  Both models fit the data almost equally well. The dashed curves show blackbody spectra corresponding to planetary brightness temperatures of 1670 K and 1514 K, the observed brightness temperatures in the {\em Spitzer} IRAC channels 1 and 2, respectively.}
\label{fig:atmmodel}
\end{figure*}

\begin{figure}[h]
\centerline{
\includegraphics[width = 0.33\textwidth, clip]{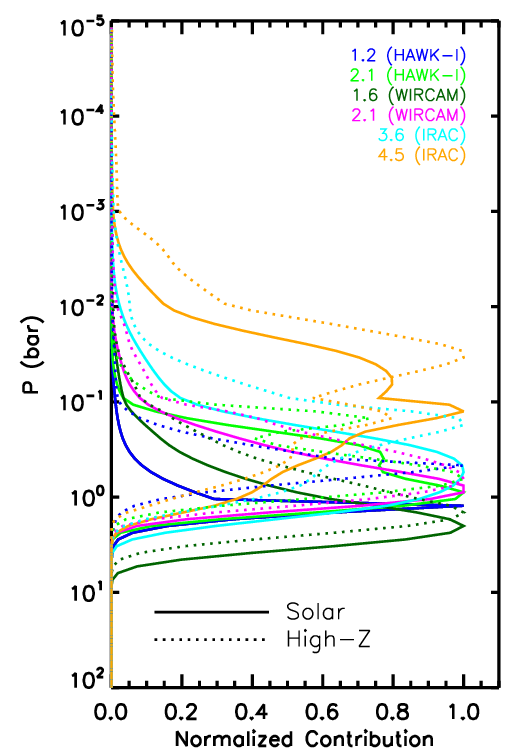}}
\caption{
Contribution functions for the atmospheric models. The solid (dashed) curves show the contribution functions for the model in Figure ~\ref{fig:atmmodel} with solar (10 \math{\times} solar, "High-\math{Z}") composition. The contribution functions are shown in all the bandpasses corresponding to the data, denoted by the central wavelengths in {\micron} and instruments in parentheses, as shown in the legend. } 
\label{fig:contr}
\end{figure}

The data rule out a strong thermal inversion in the dayside atmosphere of WASP-43b. The data and two model spectra of atmospheres without thermal inversions are shown in Figure \ref{fig:atmmodel}. The ground-based and {\em Spitzer}\/ data provide complementary constraints on the atmospheric properties. The ground-based photometric bandpasses, in narrow bands at 1.19 {\micron} and 2.09 {\micron} \citep{GillonEtal2012A&A-WASP-43b} and in broad bands at 1.6 {\micron} and 2.1 {\micron} \citep{WangEtal2013-WASP43b}, span spectral regions of low molecular opacity probe the deep layers of the atmosphere at pressures of \math{P\sim1} bar, beyond which the atmosphere is optically thick because of collision-induced opacity (the contribution functions are given in Figure\ \ref{fig:contr}). Consequently, the brightness temperatures from such ground-based data constrain the isothermal temperature structure of the deep atmosphere \citep{Madhusudhan2012}. On the other hand, the two {\em Spitzer}\/ data sets show lower brightness temperatures at 3.6 {\micron} and 4.5 {\micron} relative to the ground-based data, which is possible only if the temperature structure is decreasing outward in the atmosphere, causing molecular absorption in the {\em Spitzer}\/ bands. The presence of a strong thermal inversion, on the contrary, would have caused molecular emission leading to higher brightness temperatures in the {\em Spitzer} bands relative to the ground-based bands. Consequently, the sum total of {\em Spitzer} and ground-based data rule out a strong thermal inversion in WASP-43b's dayside photosphere. 

The molecular composition is less well constrained by the data. Several physically plausible combinations of molecules can explain the absorption in the two {\em Spitzer} bands \citep[e.g.,][]{MadhusudhanSeager2010, Madhusudhan2012}. Figure \ref{fig:atmmodel} shows two oxygen-rich models in chemical equilibrium, with C/O ratios of 0.5 (solar value) but with different metallicities (nearly solar and ten times solar) and thermal profiles, both of which explain the data almost equally well. In both cases, H\sb{2}O absorption in the 3.6 {\micron} band, and H\sb{2}O, CO, and CO\sb{2} absorption in the 4.5 {\micron} band explain the {\em Spitzer} data. The comparable fits demonstrate the degeneracy between the molecular mixing ratios (via the metallicity) and the temperature gradient. Given the current photometric data, the solar metallicity model with a steep temperature profile (green curve) produces almost as good a fit as the higher metallicity model with a shallower temperature profile (red curve). On the other hand, carbon-rich models with C/O \math{\geq}1 \citep[e.g.,][]{Madhusudhan2012}, with absorption due to CH\sb{4}, CO, C\sb{2}H\sb{2}, and HCN, could also explain the data. As such, the current data are insufficient to discriminate between O-rich and C-rich compositions. Thus, new observations are required to obtain more stringent constraints on the chemical composition of WASP-43b. Observations using the {\em HST}\/ Wide Field Camera 3 in the 1.1--1.8 {\micron} bandpass can help constrain the H\sb{2}O abundance in the atmosphere.  As shown in Figure \ref{fig:atmmodel}, an O-rich composition predicts strong absorption due to H\sb{2}O in the WFC3 bandpass, which would be absent in a carbon-rich atmosphere \citep{MadhusudhanEtal2011natWASP12batm}. Similarly, observations in other molecular bands, such as the CO band at 2.3 {\micron}, can provide constraints on the corresponding molecular mixing ratios. 

Our observations provide nominal constraints on the day--night energy redistribution fraction \citep[\math{f\sb{\rm{r}}}; ][]{MadhusudhanSeager2009} in WASP-43b. The models shown in Figure \ref{fig:atmmodel} have \math{f\sb{\rm{r}}} = 16\%--20\%, assuming zero Bond albedo. Our population of model fits to the combined {\em Spitzer} and ground-based data allow for up to \sim 35\% day--night energy redistribution in the planet. Among the acceptable models, those with higher \math{f\sb{\rm{r}}} values require cooler lower atmospheres on the dayside and hence predict lower fluxes in the ground-based channels. While such models produce an acceptable fit to all four data points overall, they predict systematically lower fluxes in the ground-based channels, some fitting the ground-based data points only at the \math{\sigma} lower error bars. Considering the ground-based points alone, without the {\em Spitzer} data, would imply a significantly higher continuum flux and correspondingly a significantly lower day--night redistribution in the planet than \sim 35\%, consistent with the findings of \citet{GillonEtal2012A&A-WASP-43b} and \citet{WangEtal2013-WASP43b}. 

The lack of a strong thermal inversion in WASP-43b is not surprising. At an equilibrium temperature of \math{\sim}1400 K, the dayside atmosphere of WASP-43b is not expected to host gaseous TiO and VO, which have been proposed to cause thermal inversions \citep{Spiegel2009, Hubeny2003, Fortney2008}, though hitherto unknown molecules that could also potentially cause such inversions cannot be ruled out \citep{Zahnle09}. The lack of a thermal inversion is also consistent with the hypothesis of \citet{KnutsonHowardIsaacson2010ApJ-CorrStarPlanet}, since the host star WASP-43 is known to be active \citep{Hellier2011-WASP43b}.

\section{CONCLUSIONS}
\label{sec:conc}

Exoplanet secondary eclipses provide us with a unique way to observe the dayside spectrum of an irradiated planetary atmosphere, where the opacities of the mixture of atmospheric trace molecules determine the thermal structure of the planetary atmosphere.

\begin{table*}[ht]
\caption{\label{table:SystemParams} 
System Parameters of WASP-43}
\atabon\strut\hfill\begin{tabular}{llc}
    \hline
    \hline
    Parameter                                                              & Value                   & Reference                 \\
    \hline
    \multicolumn{3}{c}{Eclipse Parameters}                                                                                       \\
    \hline 
    Eclipse midpoint (BJD\sb\rm{TDB}) (2011 Jul 30)                            &    2455773.3179 {\pm} 0.0003    & a                \\ 
    Eclipse midpoint (BJD\sb\rm{TDB}) (2011 Jul 29)                            &    2455772.5045 {\pm} 0.0003    & a                \\ 
    Eclipse duration  \math{t\sb{\rm 4-1}} (hr)                           &    1.25 {\pm} 0.02              & a                \\ 
    Eclipse depth {\em Spitzer} IRAC, 3.6 {\micron} (\%)                   &    0.347 {\pm} 0.013            & a                \\ 
    Eclipse depth {\em Spitzer} IRAC, 4.5 {\micron} (\%)                   &    0.382 {\pm} 0.015            & a                \\ 
    Eclipse depth VLT HAWK-I, 2.095 {\micron} (ppm)                  &    1560 {\pm} 140               & b                \\ 
    Eclipse depth VLT HAWK-I, 1.186 {\micron} (ppm)                  &    790 {\pm} 320                & b                \\ 
    Ingress/egress time \math{t\sb{\rm{2-1}}} (hr)                        &    0.264 {\pm} 0.018            & a                \\\hline  
    \multicolumn{3}{c}{Orbital parameters}                                                                                       \\ \hline
    Orbital period, \math{P} (days)                                        &    \math{P} = 0.81347436 {\pm} 1.4\tttt{-7} & c \\  
    Semimajor axis, \math{a} (AU)                                          &    0.01526 {\pm} 0.00018        & b                \\ 
    Transit time (BJD\sb\rm{TDB})                                             &    2455726.54336 {\pm} 0.00012  & b                \\ 
    Orbital eccentricity, \math{e}                                         &    0.010 $^{+0.010}_{-0.007}$   & c                \\ 
    Argument of pericenter, \math{\omega} (deg)                            &    \math{-88} $^{+5}_{-9}$      & c                \\ 
    Velocity semiamplitude, \math{K} (m\,s\sp{-1})                         &    549 {\pm} 6                  & c                \\ 
    Center-of-mass velocity \math{\gamma} (m\,s\sp{-1})                    &    \math{-3595}     {\pm} 4     & c                \\ \hline         
    \multicolumn{3}{c}{Stellar parameters}                                                                                       \\ \hline
    Spectral type                                                          &    see Section \ref{intro}       & b                \\ 
    Mass, \math{M\sb{\rm *}} (\math{M\sb{\odot}})                          &    0.717 {\pm} 0.025             & b                \\ 
    Radius, \math{R\sb{\rm *}} (\math{R\sb{\odot}})                        &    0.667 $^{+0.011}_{-0.010}$    & b                \\ 
    Mean density, \math{\rho\sb{\rm *}} (\math{\rho\sb{\odot}})            &    2.410 $^{+0.079}_{-0.075}$    & b                \\ 
    Effective temperature, \math{T\sb{\rm eff}} (K)                        &    4520 {\pm} 120                & b                \\ 
    Surface gravity, log \math{g\sb{\rm *}} (cgs)                          &    4.645 $^{+0.011}_{-0.010}$    & b                \\ 
    Projected rotation rate, \math{v\sb{\rm *} \sin(i)} (kms\sp{-1})       &    4.0 {\pm} 0.4                 & b                \\ 
    Metallicity [Fe/H] (dex)                                               &    -0.01 {\pm} 0.12              & b                \\ 
    Distance (pc)                                                          &    80 {\pm} 20                   & d                \\ \hline 
    \multicolumn{3}{c}{Planetary parameters }                                                                                    \\ \hline
    Mass, \math{M\sb{\rm p}} (\math{M\sb{\rm J}})                          &    2.034 $^{+0.052}_{-0.051}$    & b               \\ 
    Radius, \math{R\sb{\rm p}} (\math{R\sb{\rm J}})                        &    1.036 {\pm} 0.019             & b               \\ 
    Surface gravity, log \math{g\sb{\rm p}} (cgs)                          &    3.672 $^{+0.013}_{-0.012}$    & b               \\ 
    Mean density, \math{{\rho}\sb{\rm p}} (g\,cm\sp{-3})                   &    1.377 $^{+0.063}_{-0.059}$    & b               \\ 
    Equilibrium temperature (A = 0), \math{T\sb{\rm eq}} (K)                 &    1440 $^{+40}_{-39}$           & b               \\ 
    \hline
    \multicolumn{2}{l}{a - this work (parameters derived using joint fit, see Section \ref{sec:joint}).}\\
    \multicolumn{2}{l}{b      - \citet{GillonEtal2012A&A-WASP-43b}.}\\
    \multicolumn{2}{l}{c - this work (see Section \ref{sec:orbit}).}\\
    \multicolumn{2}{l}{d      - \citet{Hellier2011-WASP43b}.}
\end{tabular}\hfill\strut\ataboff
\end{table*}

WASP-43b has a 0.81 day period, making it one of the shortest-period transiting planets. It has a small semimajor axis \citep[0.01526 {\pm} 0.00018 AU,][]{GillonEtal2012A&A-WASP-43b}. WASP-43 is a low-mass star \citep[\math{M\sb{*}} = 0.717 {\pm} 0.025 \math{M\sb{\odot}},][]{GillonEtal2012A&A-WASP-43b} and is also one of the coldest of all stars hosting hot Jupiters. The close proximity of the planet probably induces large tidal bulges on the planet's surface \citep{ragozzine:2009}. The planet's projected lifetime is also unusually short for such a late-type host star, owing to tidal in-spiral. The estimated lifetime for this planet is perhaps 10 Myr--1 Gyr \citep{Hellier2011-WASP43b}.

In this paper we report two {\em Spitzer} secondary eclipse observations, using the IRAC 3.6 and 4.5 {\microns} channels. The S/N of 26 in channel 1 and 24 in channel 2 allowed a nonambiguous analysis. The final eclipse depths from our joint-fit models are 0.347\% {\pm} 0.013\% and 0.382\% {\pm} 0.015\%, in channels 1 and 2, respectively. The corresponding brightness temperatures are 1670 {\pm} 23 K and 1514 {\pm} 25 K.

Our secondary eclipse timings, along with the available RV data and transit photometry from the literature and amateur observations, provide better constraints on the orbital parameters. WASP-43b's orbital period is improved by a factor of three (\math{P} = 0.81347436 {\pm} 1.4\tttt{-7} days). The timing of our secondary eclipse observations is consistent with and suggestive of a circular orbit. 

We combined our {\em Spitzer} eclipse depths with ground-based data in the near-infrared from \citet{GillonEtal2012A&A-WASP-43b} and \citet{WangEtal2013-WASP43b} to constrain the atmospheric properties of WASP-43b. The data rule out a strong thermal inversion in the dayside atmosphere. This is particularly evident because the brightness temperatures in both the {\em Spitzer} channels are lower than those observed in the ground-based channels, suggesting temperatures decreasing outward. The data do not suggest very efficient day--night energy redistribution in the planet, consistent with previous studies, though models with up to \sim 35\% redistribution can explain the data reasonably well. Current data are insufficient to provide stringent constraints on the chemical composition.

WASP-43b is a promising planet for a variety of future observations.
Its high eclipse S/N makes it a prime candidate for dayside mapping
using eclipse ingress and egress data \citep[e.g.,][]{deWitEtal2012aapFacemap, MajeauEtal2012Facemap}. Observations in the HST WFC3 bandpass could break the degeneracy between
O-rich and C-rich atmospheric models.  Finally, the possibility of
measuring orbital decay in the future is exciting because of the unique
constraints this could place on stellar interior parameters.

\acknowledgments

We thank the observers listed in Table \ref{tab:ttv} for allowing us to
use their results and the organizers of the Exoplanet Transit
Database for coordinating the collection and uniform analysis
of these data.
We also thank contributors to SciPy, Matplotlib, the Python Programming
Language, the free and open-source community, the
NASA Astrophysics Data System, and the JPL Solar System Dynamics group
for free software and services.
The IRAC data are based on observations made
with the {\em Spitzer Space Telescope}, which is operated by the
Jet Propulsion Laboratory, California Institute of Technology,
under a contract with NASA. Support for this work was
provided by NASA through an award issued by JPL/Caltech; through the
Science Mission Directorate's Planetary Atmospheres Program, grant
NNX12AI69G; Astrophysics Data Analysis Program, grant NNX13AF38G; and by NASA Headquarters under the NASA Earth and Space Science Fellowship Program, grant NNX12AL83H. NM acknowledges support from the Yale Center for Astronomy and Astrophysics through the YCAA postdoctoral Fellowship.

{\em Facility:} Spitzer

\begin{appendix}

\section{System Parameters}
\label{sec:app}

Table \ref{table:SystemParams} lists WASP-43 system parameters derived from our analysis and the literature.

\end{appendix}

\bibliography{wasp43b}

\end{document}